\documentclass[prl,aps,onecolumn,showpacs,preprintnumbers,amsmath,amssymb,superscriptaddress]{revtex4}
\usepackage{graphicx}
\usepackage{graphicx}
\usepackage{dcolumn}
\usepackage{bm}
\usepackage{epstopdf}
\usepackage{color}

\newcommand{\brokt}[3]{\left\langle #1 \right| #2 \left| #3\right\rangle} 

\begin{document}


\title{Racah materials: role of atomic multiplets in intermediate valence systems}

\author{A. B. Shick}
\affiliation{Institute of Physics, ASCR, Na Slovance 2, CZ-18221
Prague, Czech Republic}

\author{L. Havela}
\affiliation{Department of Condensed Matter Physics,  Charles University, Ke Karlovu 5, CZ-12116, Prague, Czech Republic}

\author{A. I. Lichtenstein}
\affiliation{University of Hamburg, Jungiusstrasse 9, 20355 Hamburg,
Germany}
\affiliation{Theoretical
Physics and Applied Mathematics Department,
Ural Federal University, Mira Str.19,  620002,
Ekaterinburg, Russia}

\author{M. I. Katsnelson}
\affiliation{Radboud University Nijmegen, Heyendaalseweg 135, 6525
AJ Nijmegen, The Netherlands}
\affiliation{Theoretical
Physics and Applied Mathematics Department,
Ural Federal University, Mira Str.19,  620002,
Ekaterinburg, Russia}

\begin{abstract}
\textbf{
We address the long-standing mystery of the nonmagnetic insulating state of  the intermediate valence compound SmB$_6$.
Within a combination of the  local density approximation (LDA) and an exact diagonalization (ED) of
an effective discrete Anderson impurity model, the intermediate valence ground state
with the $f$-shell occupation $\langle n_{4f} \rangle=5.6$
is found for the Sm atom in SmB$_6$.
This  ground state is a singlet, and the first excited triplet state $\sim 3$ meV higher in the energy.
SmB$_6$ is a narrow band insulator already in LDA, with the direct band gap of $\sim 10$ meV.
The electron correlations increase the band gap  which now becomes indirect. Thus,
the many-body effects are relevant to form the indirect band gap, crucial for the idea of  ``topological Kondo insulator"  in SmB$_6$.
Also, an actinide analog PuB$_6$ is considered, and the intermediate valence singlet ground state is found for the Pu atom.
We propose that [Sm,Pu]B$_6$  belong to a new class of the intermediate valence materials with the multi-orbital
``Kondo-like" singlet ground-state. Crucial role of complex spin-orbital $f^n$-$f^{n+1}$ multiplet structure differently hybridized with ligand states
in such Racah materials is discussed.}

\end{abstract}
\date{\today}
\maketitle

Valence fluctuations in the $f$-electron based materials near the localization  threshold
attract significant attention in the condensed matter physics.
The intermediate valence has been considered originally to describe some of the rare-earth
compounds with Ce, Sm, Eu, Tm, and Yb elements. The original idea 
was that the single-particle ``promotion energy'' from 4$f$ to 5$d$ states changes the sign in these systems \cite{Khomskii, Lawrence}.
Soon, it was realized that the situation is different for the special case of Ce.
In mixed-valence Ce compounds  there is a partial delocalization of 4$f$ electrons due to direct overlap of their wave functions (4$f$ band formation),
rather than their promotion to 5$d$ band ~\cite{Johansson1975}. Later it was suggested that similar physics is relevant for 5$f$ electrons
in Pu~\cite{KatsnelsonJETPL92}.

A careful examination of various intermediate valence systems uncovers many differences between them. At first,
what are the properties of competing configurations? For Ce, this is $f^0$ and $f^1$; for Yb (like in YbB$_{12}$~\cite{Lawrence}
or elemental Yb under pressure~\cite{Colarieti-Tosti}) this is $f^{13}$ and $f^{14}$.  In both these cases one of those configurations is trivial in a many-body sense (completely empty or completely occupied 4$f$ shell).
For Sm the competing configurations are $f^5$ and $f^6$, and for Eu - 
 $f^6$ and $f^7$. In this situation, the atomic $f^n$ spin-orbital coupling (SOC) and term effects are essential. Rather than to assume the promotion between single-particle $f$- and $d$-states, one needs to consider the competition of ground-state multiplets corresponding to those configurations.  Roughly speaking, this is the case when  the Hubbard bands originated from  these multiplets are well separated. Namely, one of the sub-bands has a well-pronounced multiplet structure in solids,  and for another part of the spectrum, the multiplets are merged into a single quasiparticle sub-band~\cite{hanzawa1998}. This picture bears close similarities  to the case of $\delta$-Pu, which was called {\it ``Racah metal''}~\cite{shick13}.

Here we apply this concept to another 4$f$ and 5$f$ systems, using SmB$_6$ and PuB$_6$ as examples of  the {\it  ``Racah materials''}.
Recently, these materials were proposed as candidates to 3D topological insulators~\cite{dzero2010, Lu2013, Deng2013}, as well as ytterbium borides \cite{Weng2014}.
We  primarily focus not  on the topological properties of electronic bands in SmB$_6$ and PuB$_6$, but on the physics of valence fluctuations and
multiplet transitions in these systems. The formation of mixed valence singlet non-magnetic states in effective Anderson impurity model for these compounds crucially depends on hybridization parameters with the ligand bath orbitals and is not the universal property of such ``Kondo insulators''. Empirically, all known mixed valence Sm and Eu compounds are nonmagnetic, similar to Yb mixed-valence compounds and contrary to Tm ones \cite{Khomskii, Lawrence}; the case of Tm is special in a sense that the  ground-state multiplets for both competing configurations, $f^{12}$ and $f^{13}$ are magnetic. One can speculate that there is a general reason that mixed valence systems cannot be magnetically ordered if one of the competing ground states are nonmagnetic. We show that this is, rather, a ``play of numbers''; 
and requires the optimal hybridisation strength. In particular, we have demonstrated that a typical energy of magnetic excitations is an order of magnitude smaller than a typical energy of valence fluctuations. 

Although PuB$_6$ has lately attracted the theoretical attention, very little is known about its properties. The CaB$_6$ structure type corresponds to the cubic CsCl-type lattice in which the B$_6$ octahedra occupy the Cl site. In this structure, the B$_6$ octahedra are linked together in all six orthogonal directions and the Pu-Pu contact distance of 4.11 $ \mathrm{\AA}$  is essentially non-bonding.  The paper~\cite{smith1982} mentions only a weak temperature dependence of magnetic susceptibility. This  would suggest that the 5$f$ occupancy should be at least 5.2 or higher, as Pu systems with lower 5$f$ count are known to be magnetic~\cite{Shim2007}. It is interesting that the suggestion that PuB$_6$ has a valency lower than 3+ appeared already in the work of Smith and Fisk~\cite{smith1982} on the basis of volume and color and that the Kondo effect was considered to be responsible for the lack of magnetic moments. SmB$_6$ belongs to canonical valence fluctuation materials (valence estimated as 2.5-2.6) with the Fermi level in a hybridization gap~\cite{ Wachter}.  Careful photoemission experiments  \cite{Denlinger2013a,Denlinger2013b,Jiang2013} clearly support the complicated mixed valence nature of this ``topological insulator''.

Our aim  is to apply the state-of-the-art many-body method to develop a complete quantitative theory of electronic structure in SmB$_6$ and PuB$_6$.
We follow the "LDA++" methodology~\cite{A.I.Lichtenstein1998}, and consider
the multi-band Hubbard
Hamiltonian $H = H^0 + H^{\rm int} $,
$H^0 = \sum_{i,j,\gamma}  H^0_{i
\gamma_1, j \gamma_2}
                 c^{\dagger}_{i \gamma_1} c_{j \gamma_2}$,
                 where $i,j$ label lattice sites and
$\gamma = (l m \sigma)$  mark spinorbitals $\{ \phi_{\gamma}
\}$,
is the one-particle Hamiltonian found from \textit{ab initio} electronic
structure calculations of a periodic crystal; $H^{\rm int}$ is the
on-site Coulomb interaction~\cite{A.I.Lichtenstein1998} describing the $f$-
electron correlation.
The effects of the interaction Hamiltonian
$H^{\rm int}$ on the electronic structure are described by a
${\bf k}$-independent one-particle
self energy, $\Sigma(z)$ (where $z$ is a (complex) energy),
which is constructed with
the aid of an auxiliary impurity model describing the complete
seven-orbital 5$f$ shell. This multi-orbital  impurity model includes the full spherically symmetric
Coulomb interaction, the spin-orbit coupling (SOC), and the crystal field
(CF). The corresponding Hamiltonian can be written as \cite{Hewson}
\begin{eqnarray}
\label{eq:hamilt}
H_{\rm imp}  = & \sum_{k m m'  \sigma \sigma'}
 [\epsilon^{k}]_{m m'}^{\sigma \; \; \sigma'} b^{\dagger}_{km\sigma}b_{km'\sigma'}
 +\sum_{m\sigma} \epsilon_f f^{\dagger}_{m \sigma}f_{m \sigma}
\nonumber \\
& + \sum_{mm'\sigma\sigma'} \bigl[\xi {\bf l}\cdot{\bf s}
  + \Delta_{\rm CF}\bigr]_{m m'}^{\sigma \; \; \sigma'}
  f_{m \sigma}^{\dagger}f_{m' \sigma'}
\nonumber \\
& +  \sum_{k m m' \sigma \sigma'}   \Bigl(
[V^{k}]_{m m'}^{\sigma \; \; \sigma'}
 f^{\dagger}_{m \sigma} b_{k m'  \sigma'} + {h.c.}
  \Bigr)
\\
& + \frac{1}{2} \sum_{m m' m''  m''' \sigma \sigma'}
  U_{m m' m'' m'''} f^{\dagger}_{m\sigma} f^{\dagger}_{m' \sigma'}
  f_{m'''\sigma'} f_{m'' \sigma},
\nonumber
\end{eqnarray}
where $f^{\dagger}_{m \sigma}$ creates an electron in the 5$f$ shell and
$b^{\dagger}_{m\sigma}$ creates an electron in the ``bath''
that consists of those host-band states that hybridize with the
impurity 5$f$ shell.
The energy position $\epsilon_f$ of the impurity level, and the bath
energies $\epsilon^{k}$ are measured from the chemical potential $\mu$.
The parameters $\xi$ and $\Delta_{\rm CF}$ specify
the strength of the SOC and the magnitude of the crystal field (CF) at the impurity.
The parameter matrices  $V^{k}$ describe the hybridization between the
$f$ states and the bath orbitals at energy $\epsilon^{k}$.

The band Lanczos method~\cite{J.Kolorenc2012} is employed to
find the lowest-lying eigenstates of the many-body Hamiltonian
$H_{\rm imp}$ and to calculate the one-particle Green's function $[G_{\rm imp}(z)]_{m m'}^{\sigma \; \; \sigma'}$
in the subspace of the $f$ orbitals at low temperature
($k_{\rm B}T=1/500$ eV). The
selfenergy $[\Sigma (z)]_{m m'}^{\sigma \; \; \sigma'}$ is then
obtained from the inverse of the Green's function matrix
$[G_{\rm imp}]$.

Once the selfenergy is known, the local Green's function $G(z)$ for
the electrons in the solid,
\begin{equation}
[G(z)]_{\gamma_1 \gamma_2} = \frac{1}{V_{\rm BZ}}
\int_{\rm BZ}{\rm d}^3 k \,\bigl[z+\mu-H_{\rm LDA}({\bf
k})-\Sigma(z)\bigr]^{-1}_{\gamma_1 \gamma_2}\,, \label{eq:gf}
\end{equation}
is calculated in a single-site approximation as given in~\cite{shick09}.
Then, with the aid of the local Green's
function $G(z)$, we evaluate
the occupation matrix
$n_{\gamma_1 \gamma_2} = -\frac1{\pi}\,\mathop{\rm Im}
\int_{-\infty}^{E_{\rm{F}}} {\rm d} z \, [G(z)]_{\gamma_1 \gamma_2}$.
The matrix $n_{\gamma_1 \gamma_2}$ is used to construct an effective LDA+$U$
potential ${V}_{U}$, which is inserted into Kohn--Sham-like
equations:
\begin{equation}
[ -\nabla^{2} + V_{\rm LDA}(\mathbf{r}) + V_{U} + \xi ({\bf l} \cdot
{\bf s}) ]  \Phi_{\bf k}^b({\bf r}) = \epsilon_{\bf k}^b \Phi_{\bf
k}^b({\bf r}).
\label{eq:kohn_sham}
\end{equation}

These equations are iteratively solved until self-consistency over
the charge density is reached. In each iteration, a new Green's function
${G}_{\mathrm{LDA}}(z)$ (which corresponds to $G(z)$ from Eq.(\ref{eq:gf})
with the self energy $\Sigma$ set to zero), and a new value of the 5$f$-shell
occupation are obtained from the solution of Eq.~(\ref{eq:kohn_sham}). Subsequently,
a new self energy $\Sigma(z)$ corresponding to the updated $f$-shell occupation is
constructed. Finally, the next iteration is started by evaluating the new
local Green's function,~Eq.(\ref{eq:gf}).

SmB$_6$ and PuB$_6$ crystalize in the CaB$_6$-structure with the space group $Pn3m$ (221), as shown in Fig.~S1 (supplementary information).
The experimental lattice constants of 4.1333 $ \mathrm{\AA}$ for SmB$_6$
and  4.1132  $ \mathrm{\AA}$ for PuB$_6$ are used.
In the calculations we used  an in-house
implementation~\cite{shick99,shick01} of the FP-LAPW method
that includes both
scalar-relativistic and spin-orbit coupling effects. For SmB$_6$, the Slater
integrals were chosen as $F_0=6.87$~eV, and $F_2=9.06$ eV, $F_4=6.05$
eV, and $F_6=$ 4.48 eV~\cite{vdMarel1988}. They corresponds to commonly
accepted values for Coulomb~$U=6.87$ eV and Hund exchange~$J = 0.76$ eV, and
are in the ballpark of the parameters commonly used in the calculations of the
rare-earth materials~\cite{Lebegue2006}. For PuB$_6$, the Slater
integrals  $F_0=4.0$~eV, and $F_2=7.76$ eV, $F_4=5.05$
eV, and $F_6=$ 3.07 eV were chosen~\cite{KMoore2009}. They corresponds to commonly
accepted values for Coulomb~$U=4.0$ eV and exchange~$J = 0.64$ eV.
The SOC parameters $\xi=0.16$ eV for SmB$_6$,  and
$0.29$ eV for PuB$_6$  were determined from LDA calculations.
CF effects were neglected and
$\Delta_{\rm CF}$ was set to zero.
For the double-counting term entering the definition of the
LDA+$U$ potential, $V_U$, we have adopted the fully-localized (or
atomic-like) limit (FLL)  $V_{dc} = U (n_f-1/2) - J(n_f-1)/2$.
Furthermore, we set the radii of the atomic spheres
to 2.85~a.u.~(Sm), 3.0~a.u.~(Pu), 1.53~a.u.~(B).
The parameter $R_{Sm} \times K_{max}=9.98$ determined the
basis set size, and the Brillouin zone (BZ) sampling was performed
with 1331 $k$~points. The self-consistent procedure defined by
Eqs.~\ref{eq:hamilt}--\ref{eq:kohn_sham} was repeated until the
convergence of  the $f$-manifold occupation $n_f$ was better than
0.01.

\begin{figure}[htbp]
\centerline{\includegraphics[angle=0,width=0.5\columnwidth]{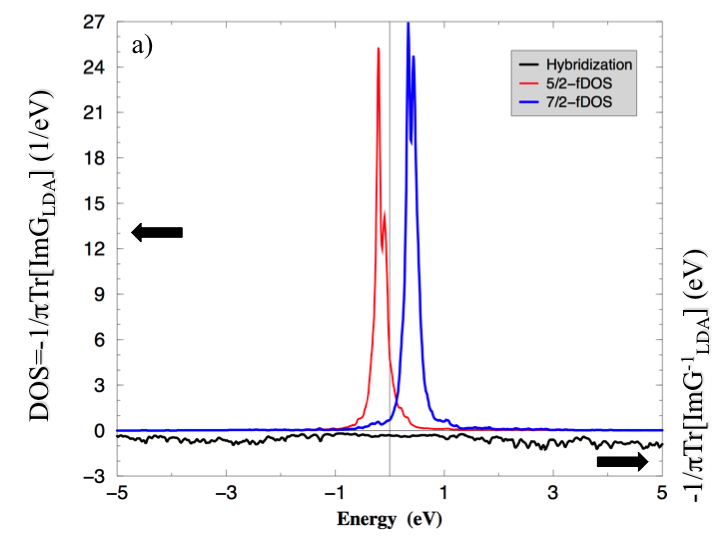}
\includegraphics[angle=0,width=0.5\columnwidth]{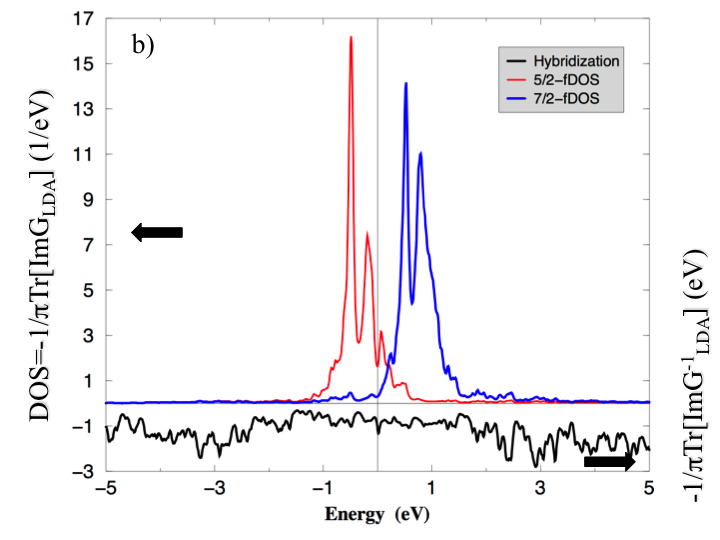}}
\caption{LDA $j$ = 5/2, 7/2 projected DOS, and LDA
hybridization function ${{\Delta}(\epsilon) \over \pi} = - {1 \over {\pi}} Im Tr
[G^{-1}(\epsilon + i \delta)]$ for SmB$_6$ (a) and PuB$_6$(b) .} \label{hybridization}
\end{figure}

In order to determine the bath parameters $V^{k}$ and $\epsilon^{k}$, we assume that the LDA
represents the non-interacting model. We then associate the LDA Green's
function ${G}_{\mathrm{LDA}}(z)$ with the Hamiltonian of
Eq.~(\ref{eq:hamilt}) when the coefficients of the Coulomb interaction matrix are set to zero
($U_{mm'm''m'''}=0$). The hybridization function ${\Delta(\epsilon)}$ is then estimated as ${\Delta}(\epsilon) =
- \mathop{\rm Im}\mathop{\rm Tr} [G^{-1}_{\rm LDA}(\epsilon + i \delta)]$.
The curve obtained for $ {1 \over \pi} {\Delta}(\epsilon)$ is shown in
Fig.~\ref{hybridization}, together with the LDA
density of states (total and $j=5/2,7/2$-projected).
The results  show that the hybridization matrix
is, to a good approximation, diagonal in the $\{j,j_z\}$ representation.
Thus, we assume the first and fourth terms in the impurity model,
Eq.~(\ref{eq:hamilt}), to be diagonal in $\{j,j_z\}$, so that we only need to specify
one bath state (six orbitals) with $\epsilon^{k=1}_{j=5/2}$ and $V^{k=1}_{j=5/2}$, and
another bath state (eight orbitals) with $\epsilon^{k=1}_{j=7/2}$
and $V^{k=1}_{j=7/2}$. Assuming that the most important
hybridization is the one occurring in the vicinity of $E_F$, as suggested by the curve shown in
Fig.~\ref{hybridization}, the numerical values of the bath parameters $V^{k=1}_{5/2,7/2}$
are found from the relation~\cite{Gunnarsson89}
$ \pi \sum_{k} {|V_{k}^{j}|}^2 \delta(\epsilon_{k}^{j} - \epsilon) = 
- \Delta(\epsilon)/N_f$ averaged over the energy interval,
$E_F - 0.5$ eV $\le \epsilon \le E_F + 0.5$ eV,
with $N_f=6$ for $j=5/2$ and $N_f=8$ for $j=7/2$.
The bath-state energies
$\epsilon^{k=1}_{5/2,7/2}$ shown in Table~\ref{parameters} are adjusted to approximately reproduce
the LDA $f$-states occupations $n_f^{5/2}$ and $n_f^{7/2}$.

The magnitude of $\Delta(E_F)$ ($\equiv \Delta_v$) is a characteristic energy of the valence fluctuations, in a sense that for the time scale
$t  >  {\hbar \over \Delta_v}  \equiv  \tau_{fl}$
the system behaves as a homogeneous with the physical properties
which are intermediate between those for Sm$^{2+}$ and  Sm$^{3+}$ whereas for  $t  < \tau_{fl}$  it is a random
configuration of "frozen"  Sm$^{2+}$ and  Sm$^{3+}$ ions~\cite{Khomskii, Lawrence, Colarieti-Tosti}. 
Lattice parameter and core-level X-ray spectra serve as examples of the properties of the first and the second kind.

\begin{table}[htbp]
\caption{$f$-states occupations  $n_f^{5/2}$ and $n_f^{7/2}$, and bath state parameters
 $\epsilon^1_{5/2,7/2}$  (eV),
$V^{1}_{5/2,7/2}$ (eV) for  Sm and Pu-atoms in SmB$_6$, and
PuB$_6$ from LDA calculations.} \label{occup}
\begin{center}
\begin{tabular}{lccccccc}
\hline
 Material & $n_f^{5/2}$ &$n_f^{7/2}$&$\epsilon_1^{5/2}$&$V_{1}^{5/2}$&$\epsilon_1^{7/2}$&$V_{1}^{7/2}$   \\
\hline
 SmB$_6$   & 5.28 & 0.26 & -0.20 & 0.16   & 0.07 &0.15\\
 PuB$_6$ & 4.89 & 0.40 & 0.13 & 0.26   & -0.05 &0.17\\
\hline
\end{tabular}
\end{center}
\label{parameters}
\end{table}

\subsection{SmB$_6$}
First, we focus  on Sm$B_6$, and discuss the solution of Eq.(\ref{eq:hamilt}).
The ground state of the cluster formed by the
4$f$~shell and the bath is given by a non-magnetic singlet with all angular moments of the 5$f$-bath cluster equal to zero
($S = L = J =0$). For the 4$f$ shell alone, the $\langle n_f \rangle=5.63$, and the $\langle n_{bath} \rangle=6.37$  bath states.
Note that  $\langle n_f \rangle$ slightly exceeds its LDA value of  $5.54$.
The expectation values of the
spin $S_f$, orbital $L_f$ and total $J_f$ angular moments can be calculated as $\langle \hat X_f^2 \rangle=X_f(X_f+1)$
($X_f = S_f, L_f, J_f$), giving  $S_f = 2.77$, $L_f = 3.80$, and $J_f = 1.88$. The ground state is separated from the
first excited state by the gap $\Delta_m =$2.6 meV. Surprisingly, this value is in a very good agreement with the experimental  activation gap
value of 3 meV~\cite{robler2014}.  
{This gap should show itself in the magnetic susceptibility, which is anticipated to behave as $1/[T+T_m]$ at high temperatures, with saturation below $T_m$  temperature  $\sim \Delta_m$, in qualitative agreement with the experimental data~\cite{Wachter}, and other experiments which measure the two-particle excitations.}
This excitation in two-particle spectrum can be contrasted with first single-particle photoemission peak
around ~20 meV~\cite{Denlinger2013a}.  It is important to mention that formation of mixed-valance multi-orbital singlet in effective Anderson model
is very sensitive to hybridization parameters (Table~\ref{parameters}) and with relative small changes
the magnetic ground states is formed in ED calculations. 
It is also important that this magnetic exciton energy is an order of magnitude smaller than the
energy  of  the valence fluctuations $\Delta_v \approx 70$ meV. This means that the nonmagnetic character of the ground state is not directly related to 
the valence fluctuations: the system possesses local magnetic moments in the energy (and temperature) range between $\Delta_m$ and $\Delta_v$, that is,
within the homogeneous intermediate valence regime.

\begin{figure}[htbp]
\centerline{\includegraphics[angle=0,width=0.5\columnwidth]{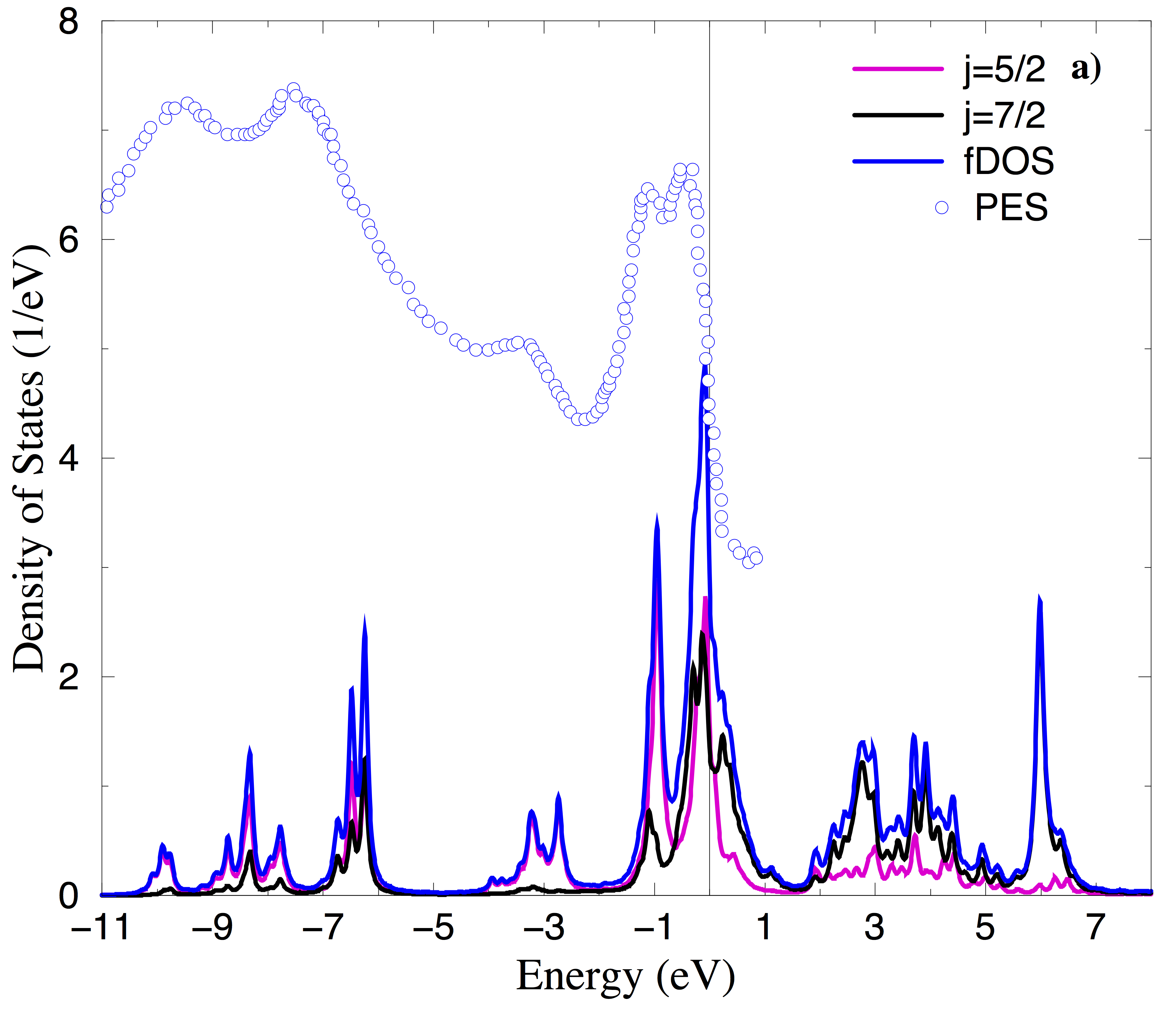}
\includegraphics[angle=0,width=0.5\columnwidth]{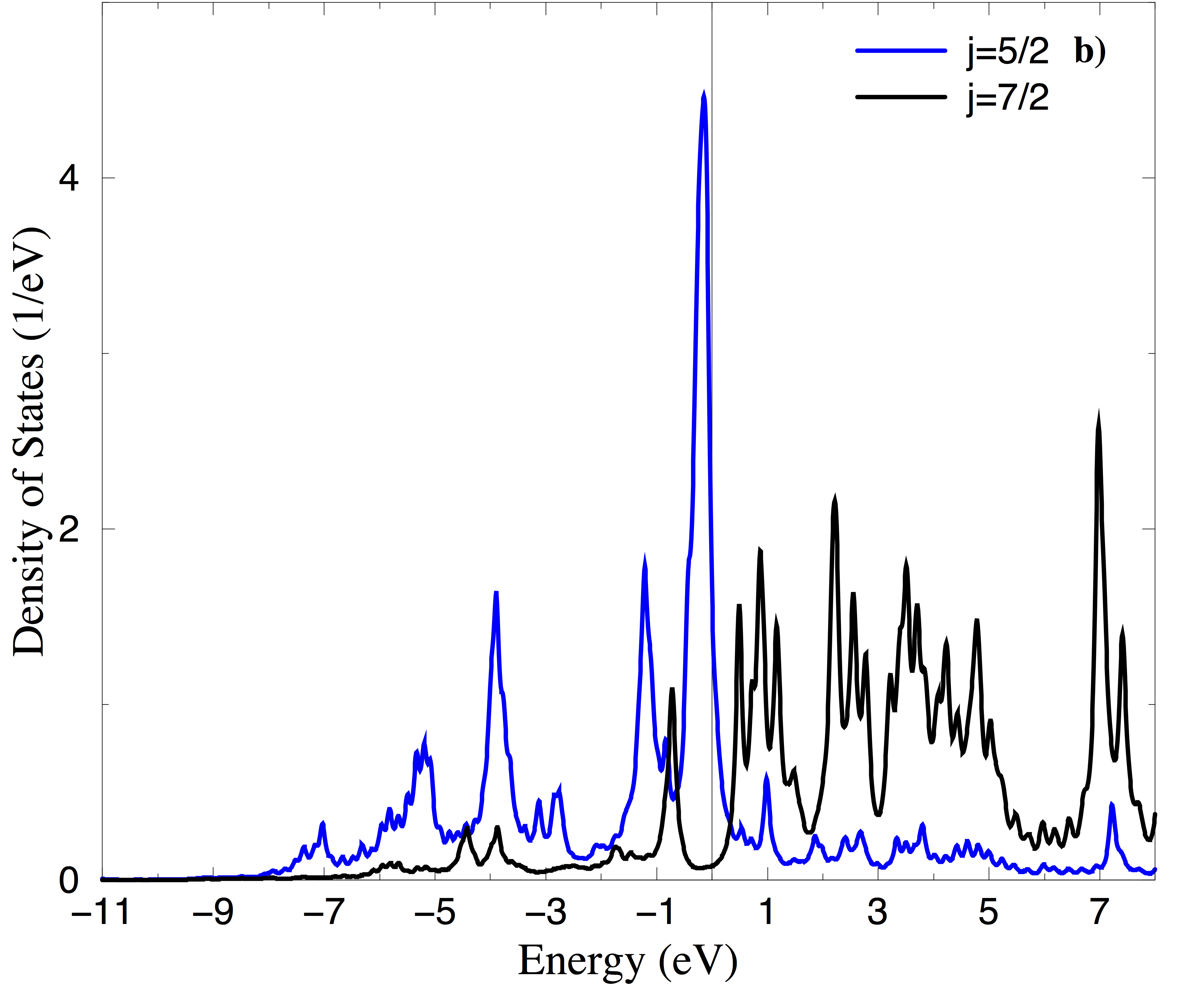}}
\caption{$f$-electron density of states (fDOS, and $j$ = 5/2, 7/2
projected) for the Sm atom in SmB$_6$ (a), and the Pu atom in PuB$_6$ (b).
Also comparison with the experimental XPS spectra is given for SmB$_6.$ }
\label{fig:dos}
\end{figure}

The $f$-orbital density of states (DOS) obtained from Eq.~(\ref{eq:gf}) for SmB$_6$ is shown in Fig.~\ref{fig:dos}(a).
The $f$-DOS is in agreement with the experimental x-ray photoelectron spectra (XPS)~\cite{chazalviel1976}, {and previously reported Hubbard-I calculations~\cite{thunstrom2009}}.
The many-body resonances near the Fermi energy are produced by $f^6 \rightarrow f^5$ multiplet transitions, they are in a way analogues to the {\it Racah} peaks,  specific
transitions between Racah  multiplets~\cite{Racah1949} of $f^n$ and $f^{n \pm 1}$.

Fig.~S2(a) (supplementary information) shows the LDA band structure together with
the band structure calculated from the solutions of Eq.~(\ref{eq:kohn_sham}), which
represents an extended LDA+U band structure
with the 5$f$-states occupation matrix obtained from the local
impurity Greens function Eq.(\ref {eq:gf}) (LDMA).
Note that the LDA band structures are very similar to  previously reported
results of WIEN2K for SmB$_6$~\cite{Lu2013}.

\begin{figure}[htbp]
\centerline{\includegraphics[angle=0,width=0.9\columnwidth]{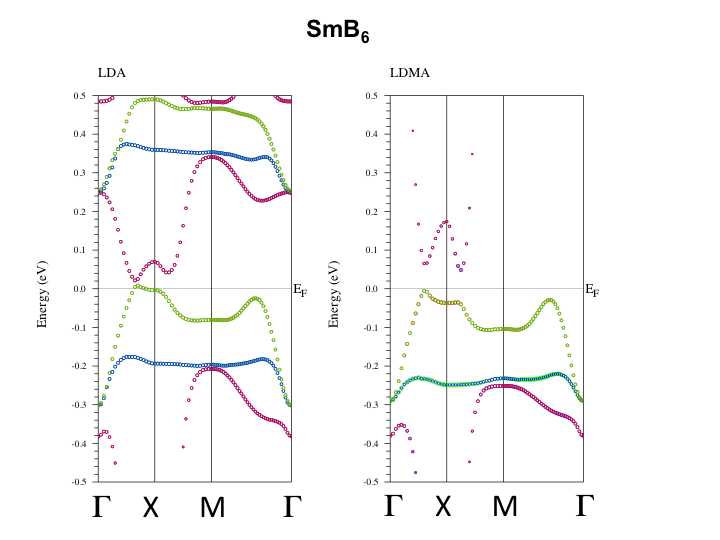}}
\centerline{\includegraphics[angle=0,width=0.9\columnwidth]{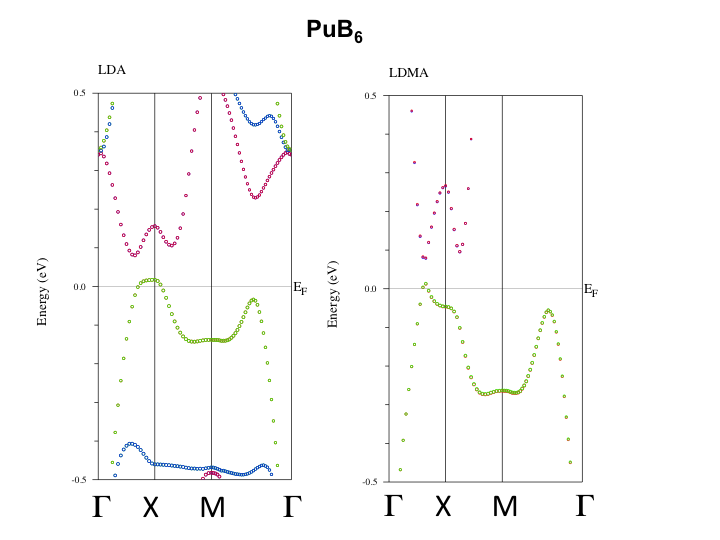}}
\caption{SmB$_6$ (a) and PuB$_6$ (b)  LDA and LDMA band structure on the small energy scale.
The circles indicate the f-character of the electronic states.}
\label{fig:bands_lda2}
\end{figure}

A more detailed look at the band structure is shown in Fig.~\ref{fig:bands_lda2}(a)
SmB$_6$ is close to a very narrow band insulator already in LDA. There is a tiny amount
of holes in the vicinity of the X-point (similar to Ref.~\cite{Lu2013}) and a direct gap of $\sim$30 meV
right above. When the Coulomb interaction is added, it becomes an indirect band insulator with
the gap of $\sim$60 meV. Note that the band-gap value exceeds somewhat the experimental
gap of  around 20 meV.  {Incorporating the dynamical self-energy effects into the LDMA band structure,
as described in the supplemental material Fig.~S3, we obtain that the indirect band gap is somewhat reduced to $\sim30$ meV 
becoming closer to the experimental value of $20$ meV.} 

It is known that the $d$-$f$ Coulomb interaction $G$ (Falicov-Kimball interaction) plays a role for the intermediate valence~\cite{Khomskii, kikoin1983, irkhinJETP86}. This interaction leads to the excitonic renormalization of the effective hybridization.
The effective hybridization $V_{eff}$ between $d$ and $f$ states with the many-body renormalization can be calculated using the electronic structure expression~\cite{irkhinJETP86} which for zero temperature reads:
\begin{equation}
V_{eff} \bigl[1 -  \; \frac{G L_{0}}{\int{\rm d} \epsilon N( \epsilon )}\bigr] = V \; , \;
L_{0} =  \int{\rm d \epsilon} \frac{N( \epsilon )}{(\epsilon^2 + 4 V_{eff}^2)^{1/2}} \; .
 \label{eq:veff}
\end{equation}
In this Eq.~(\ref{eq:veff}),  
$N( \epsilon )$
is the total $DOS$ without the $f$-projected contribution,
and $V$ is the LDA hybridization from the Table~\ref{parameters}. Importantly, the renormalized hybridization turns out to be quite strongly temperature dependent~\cite{irkhinJETP86}.

The parameter $G$ can be determined as the derivative of the center of the 5$d$ band with respect to the number $n_f$ of 4$f$ electrons~\cite{Colarieti-Tosti}). In practice, we have varied $n_f$ by changing the double-counting term from the FLL  ($n_f$=5.63)
to the ``around-mean-field'' (AMF, $n_f$=5.68), and obtained the Falicov interaction parameter of 3.8 eV. Solution of
the Eq.~(\ref{eq:veff}) yields the $V_{eff}/V$ renormalization of 1.77.

Thus, the $d$-$f$ excitonic effects enhance the hybridization making the hybridization gap larger and therefore favoring the topological insulator behavior. We performed the calculations with this renormalised
$V_{eff}$ in Eq.~(\ref{eq:hamilt}), and obtained again the singlet ground state. The  $\langle n_f \rangle=5.61$ has  decreased slightly.
This numerical stability of the Sm singlet ground state with respect to a hybridization strength is important
since experiments~\cite{altshulerJETPL84} show a strong temperature dependence of the energy gap in SmB$_6$ which cannot be explained in a purely hybridization model; they were explained in Ref.~\cite{irkhinJETP86} via excitonic effects. Recently, a strong decrease of the hybridization gap with the temperature increase in SmB$_6$ was found in ARPES~\cite{Denlinger2013b}. This can be also considered as a confirmation of strong many-body (excitonic) renormalization of the hybridization. 

\begin{figure}[!htbp]
\centerline{\includegraphics[angle=0,width=0.75\columnwidth]{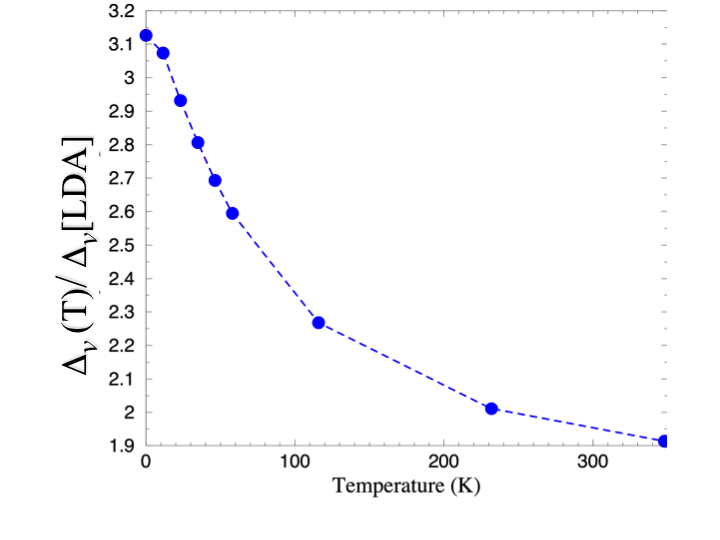}}
\caption{The temperature dependence of the hybridization gap (indirect), $\Delta_v(T)/ \Delta_v$[LDA]$= (V_{eff} (T) / V)^2$ 
calculated in the Falicov-Kimball model Eqs.~(\ref{eq:veff}), (\ref{eq:veffT}).}
\label{fig:gap}
\end{figure}

To estimate the temperature dependence of the hybridisation due to Falicov-Kimball interaction we use the theory~\cite{irkhinJETP86} for the finite 
temperatures, that is, Eq.(~\ref{eq:veff}) with the replacement,
\begin{equation}
L_{0} \rightarrow  \;
L =  \int{\rm d \epsilon} \frac{N( \epsilon )}{(\epsilon^2 + 4 V_{eff}^2)^{1/2}}  \big[ 1 - f(\epsilon_1) -  f(\epsilon_2)  \big]\; ,
 \label{eq:veffT}
\end{equation}
where $\epsilon_{1,2} = {1 \over 2} ({(\epsilon^2 + 4 V_{eff}^2)^{1/2}} \pm \epsilon)$, and $f(\epsilon)= { 1 \over e^{\epsilon / {k_B T}} + 1} $
is the Fermi function. The results are shown in Fig.~\ref{fig:gap}.

The presence of the non-magnetic $f^6$ multiplet is crucially important for the non-magnetic singlet ground state of SmB$_6$. For instance,
in the intermediate valence TmSe (competition of $f^{12}$ and $f^{13}$ configurations) the ground state is magnetic since both configurations are magnetic.
 At the same time, there is no  ``theorem'' that for the non-magnetic ground state of one of the competing configurations the system cannot be magnetic, and the
specific values of the relevant parameters are important. As we have seen, even typical energy scales for the magnetic ($\Delta_m$) and valence ($\Delta_v$)
fluctuations are different. 

\begin{figure}[!htbp]
\centerline{\includegraphics[angle=0,width=0.75\columnwidth]{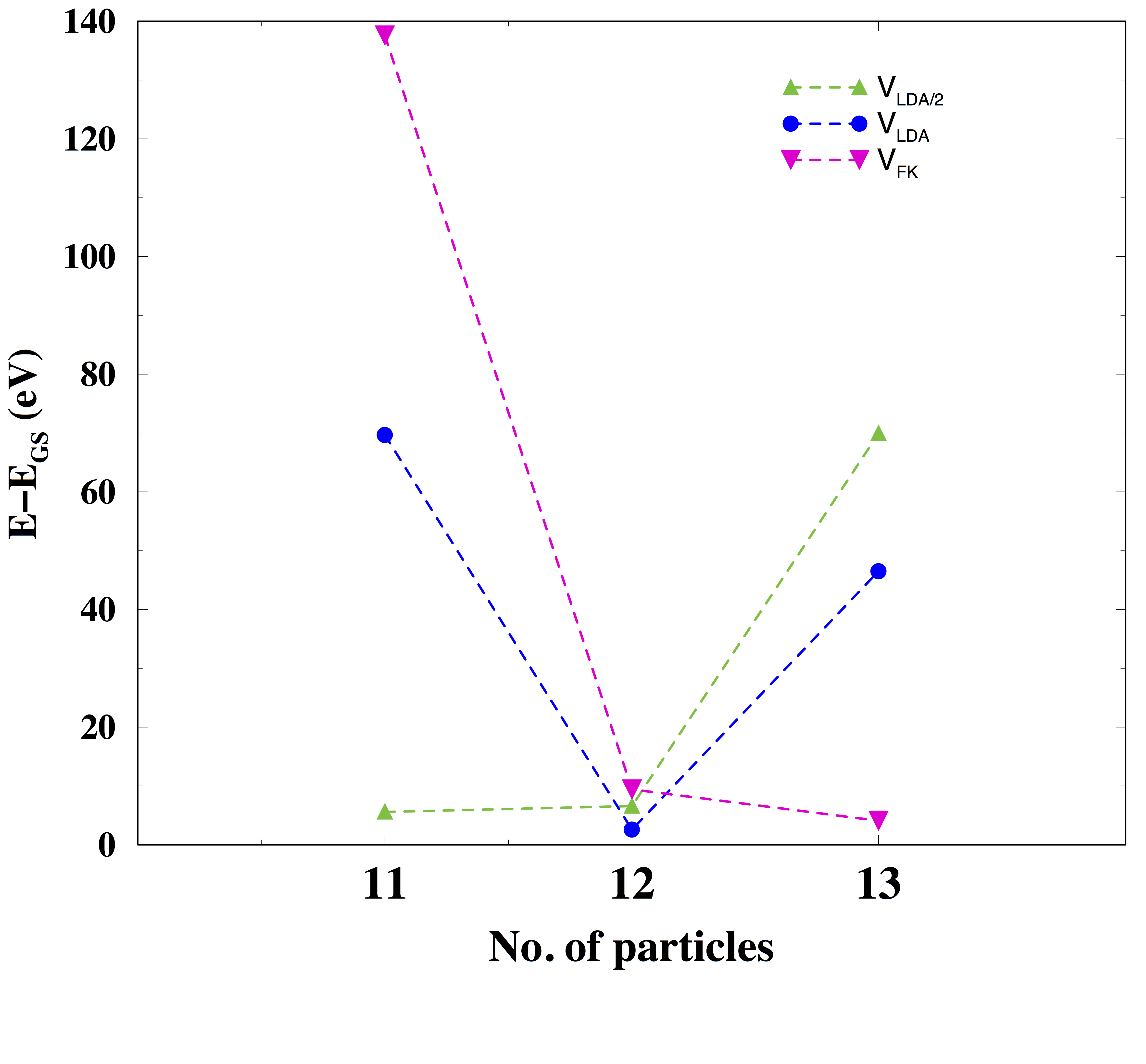}}
\caption{The energy difference between the first excited eigenstate for given number of particles ($N=n_{bath}+n_{f}$)  and the ground state of the Eq.~(\ref{eq:hamilt}) for different values of hybridization : (i)  $V_{LDA/2}$ reduced by a factor of 2 from those calculated in LDA and given in Tab.~\ref{parameters}, 
(ii) $V_{LDA/2}$ from Tab.~\ref{parameters} ; (iii) renormalised by the Falicov-Kimball model Eq.~\ref{eq:veff}.}
\label{fig:eig}
\end{figure}

For the $f$-shell occupation  $n_{f}$ of 5.6, we show in Fig.~\ref{fig:eig}  the energy difference between the first excited eigenstate for given number of particles ($N=n_{bath}+n_{f}$)  and the ground state of the Eq.~\ref{eq:hamilt} for different values of hybridization : those calculated in LDA and given in Tab.~\ref{parameters}, reduced by a factor of 2, and renormalised by the Falicov-Kimball model, as it was described above.
In all those calculations, the ground state is a non-magnetic singlet with $N=12$. For the LDA hybridization, the lowest excited state belongs to the same $N=12$, and
is lying 3 meV above the ground state. The excited magnetic $N=11$ and $N=13$ states are shifted upwards in the energy by 70 meV and 47 meV respectively.
When the hybridization is reduced (twice smaller than its LDA value), a non-magnetic ground state singlet with $N=12$ is by 6 meV lower than almost degenerate
$N=11$ and $N=12$ magnetic excited states. The $N=13$ excitation is substantially (by 70 meV) higher in the energy.  At the same time, for the hybridization 
renormalised by the Falicov-Kimball model Eq.~\ref{eq:veff}, the situation is inverse: the lowest magnetic excited state of 4 meV belongs to  $N=13$, next (9 meV) has the same  $N=12$, and the $N=11$ excitation exceeds the singlet ground state by 139 meV. Further increase of the hybridization, say by a factor of 2 with respect to the LDA value, leads to occurrence of the magnetic $N=13$ ground state.  

 In this sense, while it is possible to call the situation ``Kondo singlet with high Kondo temperature'' (which simply means a formation of singlet from the states of localized and itinerant electrons) one should keep in mind that microscopically some effects beyond the Kondo or Andreson model, such as Falicov-Kimball interactions can contribute significantly. There is an essential difference with  various Ce- and Yb-based  systems where multiplets are not important, and the situation is indeed closer to the ÒKondo lattice with high Kondo temperatureÓ.
\subsection{PuB$_6$}
Now we turn to the case of PuB$_6$.  In this case, the  hybridization strength is substantially increases
(see Table~\ref{parameters}) .  The hybridized ground state of the Pu atom in PuB$_6$, the solution
of  Eq.(\ref{eq:hamilt}),  is a non-magnetic singlet with all angular moments of the 5$f$-bath cluster equal to zero
($S = L = J =0$). It consists of  $\langle n_f \rangle=5.49$ $f$~states and
$\langle n_{bath} \rangle=8.51$  bath states. As in the case of  SmB$_6$, the magnetic moment of the 5$f$ shell
($S_f = 2.23$, $L_f = 3.68$, $J_f = 1.94$) is completely compensated by the moment carried by the electrons in the conduction band. As the value of the 5$f$ magnetic moment fluctuates in time, because of the intermediate valence electronic configuration, this compensation must be understood as dynamical in nature. The same situation is realized in $\delta$-Pu ($S_f = 2.11$, $L_f = 4.21$, $J_f= 2.62$), whose ground state is found to be
a non-magnetic singlet with $\langle n_f \rangle=5.21$ and $\langle n_{bath} \rangle=8.79$~\cite{shick13}.

The $f$-orbital density of states (DOS) obtained from Eq.~(\ref{eq:gf}) for PuB$_6$ is shown in Fig.~\ref{fig:dos}(b).
No experimental photoelectron spectra available in this case.  As in $\delta$-Pu, there are three many-body resonances near the Fermi energy which are produced by $f^6 \rightarrow f^5$  Racah multiplet transitions.

The LDA band structure is very similar to  previously reported
results of WIEN2K for PuB$_6$~\cite{Deng2013}  as shown in Fig.~S2(b)  (supplementary information), and, in more details,
in Fig.~\ref{fig:bands_lda2}(b).
Already in the LDA, PuB$_6$, is close to an insulator  with a small amount of holes near the X-point,
and the indirect band gap of $\sim$60 meV. In the LDMA,  PuB$_6$ becomes almost an insulator, with the tiny fraction of holes
near the X-point, and direct band gap of  $\sim$60 meV (see Fig.~S2(b) and Fig.~\ref{fig:bands_lda2}(b)).

As to PuB$_6$, we have very little material for comparison with experiment, as there is much less data not only comparing to rare earth borides but also with respect to other Pu compounds. A group of analogous compounds with an energy gap and non-magnetic behaviour are Pu chalcogenides PuX, with X = S, Se, Te. Photoelectron spectra~\cite{Gouder2000, Durak2004} reveal a pronounced fingerprint of the final-state 5f$^5$ multiplet close to the Fermi level, which implies that the 5f$^6$ state must contribute to the ground state. The Pu chalcogenides have also qualitatively similar non-metallic conductivity explained by hopping~\cite{ichas2001}, qualitatively analogous not only to  SmB$_6$, but also to Sm chalcogenides.

\section{Conclusions}
The  electronic structure calculations are performed within the density functional plus dynamical mean-field theory (``LDA++''~\cite{A.I.Lichtenstein1998}) approach combining the local density approximation (LDA) with an exact diagonalization (ED) of the Anderson impurity model for  SmB$_6$ and PuB$_6$. The intermediate valence singlet ground states are found for these materials.  When the Coulomb $f-f$ (Hubbard) correlations are included, SmB$_6$ becomes an indirect band gap insulator, while PuB$_6$ is a direct band gap insulator. A combined effect of specific Racah multiplet structure with intermediate valence behavior of these compounds results in complicated excitation spectrum clearly seen in different photoemission experiments. Formation of singlet ground state in the ED impurity calculations is not universal and crucially depends on structure of two mixed valance multiplets and parameters of effective Anderson model. The Coulomb $f-d$ (Falicov-Kimball) interactions increase essentially the effective hybridization influencing  additionally the singlet state. Their role may be essential in explanation of recently observed temperature-dependent electronic structure of SmB$_6$~\cite{Denlinger2013b}. 
The calculations illustrate that many-body effects are relevant to form the {\ indirect} band gap. 
In PuB$_6$ we have found also a mixed-valent singlet ground state with basically the same multiplet physics as was discussed earlier for $\delta$-Pu~\cite{shick13}. 

To emphasize the role of multiplet effects in competing valence states for this class of mixed valence systems, we suggest the term ``Racah materials''. The distinguishing feature for these materials is that part of electron excitation spectrum originated from one the valence configurations is more atomic like (with well-pronounced multiplets) whereas for the other valence configuration it is more itinerant-like.
{The consept of ``Racah materials'' is somewhat related to the idea of Òquasiparticle multipletsÓ~\cite{yee2010}. Those are represented by atomic-like multiplet transitions $f^6$-$f^5$ near the Fermi edge. In addition, there is a second part at the lower energy ($f^5$-$f^4$) which are more itinerant-like and merged into the quasi-particle subband~\cite{hanzawa1998}. Co-existence of these two types of the Hubbard bands in SmB6 and PuB6 defines them as Racah materials.} 


\section{Acknowledgements}
The support from the Czech Republic grant GACR No. 15-07172S is acknowledged. 
AIL acknowledges financial support by Grant No. DFG LI 1413/8-1.
MIK acknowledges financial support
by ERC Advanced Grant No. 338957
and by NWO via Spinoza Prize.

\section{Author contributions}
ABS, MIK and AIL conceived and supervised the project. ABS and MIK performed the 
computations. All authors contributed to the interpretation of the data and to the writing of the
manuscript.

\section{Additional information}
{\bf Competing financial interests:} The authors declare no competing financial interests.



\clearpage

\setcounter{figure}{0}
\makeatletter
\renewcommand{\thefigure}{S\@arabic\c@figure}
\makeatother

\setcounter{equation}{0}
\makeatletter
\renewcommand{\theequation}{S\@arabic\c@equation}
\makeatother

\setcounter{table}{0}
\makeatletter
\renewcommand{\thetable}{S\@arabic\c@table}
\makeatother

\makeatletter
\renewcommand{\@biblabel}[1]{[#1]}
\makeatother

\centerline{\textbf{ \em{Supplementary Information for}}}

\centerline{\textbf{Racah materials: role of atomic multiplets in intermediate valence systems}}

\renewcommand{\thefigure}{S1}
\begin{figure}[htbp]
\centerline{\includegraphics[angle=0,width=0.5\columnwidth]{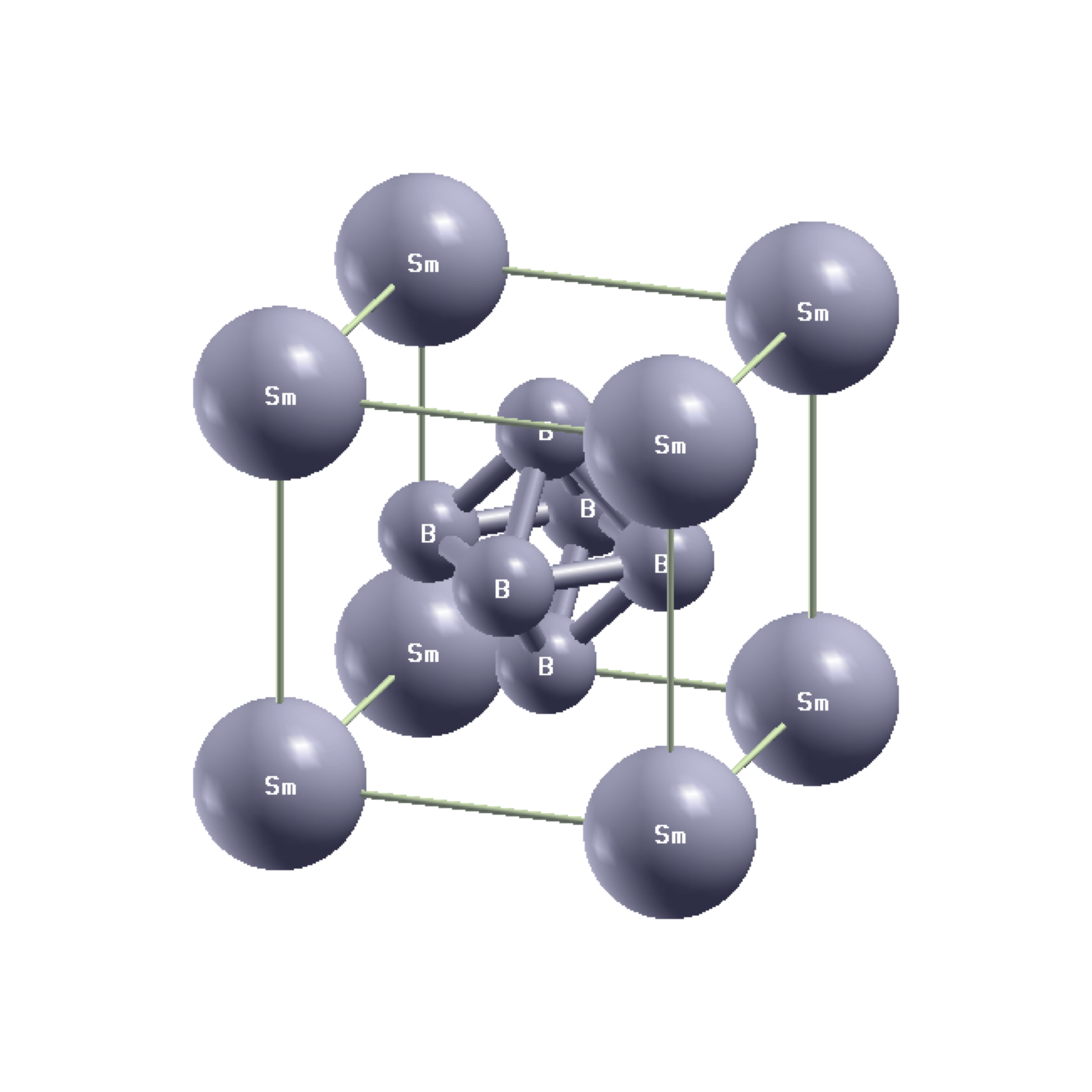}}
\caption{The crystal structure of SmB$_6$.}
\label{fig:struct}
\end{figure}

\renewcommand{\thefigure}{S2}
\begin{figure}[htbp]
\centerline{\includegraphics[angle=0,width=0.9\columnwidth]{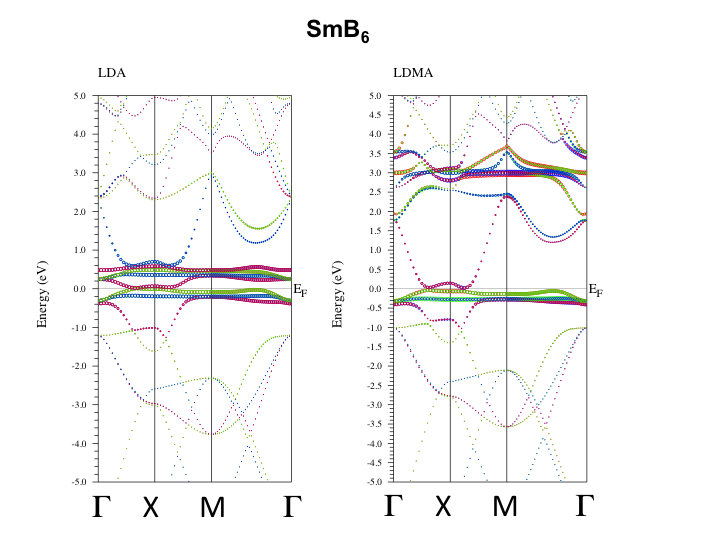}}
\centerline{\includegraphics[angle=0,width=0.9\columnwidth]{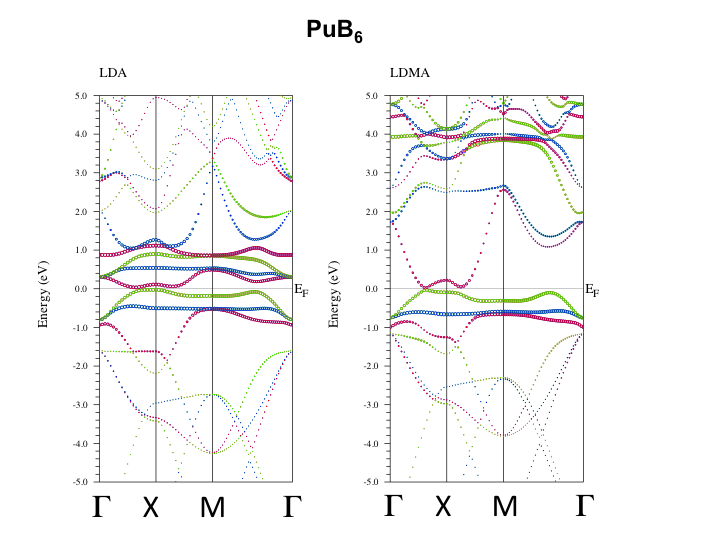}}
\caption{SmB$_6$ (top) and PuB$_6$ (bottom)  LDA and LDMA band structure. }
\label{fig:bands_lda}
\end{figure}

\newpage

In order to incorporate the dynamical self-energy effects into the LDMA band structure shown in Fig.~3 and
Fig.~S2, we make use of the first-order perturbation theory for the Green's function of the Eq. (3) with respect to 
$[\Sigma(z) \; - \; V_{U}]$, and write the $\bf k$-resolved spectral density  $A({\bf k}, z)$ as,

\begin{equation}
\label{eq:sd}
A({\bf k}, z) = - {Im \over \pi} \sum_{n} \Big[
{1 \over {z + \mu -  \epsilon^n_{\bf k}}}   \; + \;  
\frac{\brokt{\Phi^n_{\bf k}}{\Sigma(z)-V_{U}} {\Phi_{\bf k}^n}} {(z + \mu -  \epsilon^n_{\bf k})^2} 
 \Big].
\end{equation}

\renewcommand{\thefigure}{S3}
\begin{figure}[htbp]
\centerline{\includegraphics[angle=0,width=1.0\columnwidth]{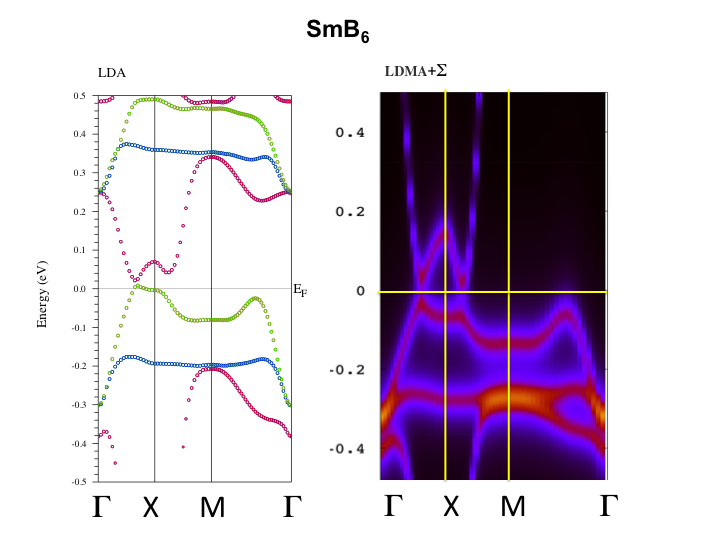}}
\caption{SmB$_6$   LDA and LDMA + $\Sigma$ band structure. }
\label{fig:arpes}
\end{figure}

We plot the spectral density in Fig.~\ref{fig:arpes} together with the LDA band structure for SmB$_6$ . We notice that the effect of $\Sigma(z)$ on the LDMA
band structure  shown in Fig.~3(top) is rather small.  The indirect band gap is somewhat reduced to $~30$ meV becoming closer to the experimental
value of $20$ meV. Also, it resembles some features of the DMFT calculated spectral DOS~\cite{Kim2014} as well as  the experimental angular resolved photoemission (ARPES)~\cite{Denlinger2014}.  Since the experimental ARPES is known be very surface sensitive, careful comparison with the theoretical
calculations including the surface calculations is needed. These studies are beyond the scope of the present work, and left for the future.

\end{document}